\documentclass{llncs}

\let\ifcomment\iftrue 

\pdfoutput=1

\usepackage{geometry}
\usepackage[utf8]{inputenc}
\usepackage[T1]{fontenc}
\usepackage{fullpage}
\usepackage{amsmath}
\usepackage{amssymb}
 \usepackage{amsthm}
 \usepackage{amsfonts}
\usepackage{physics}
\usepackage{marginnote}
\usepackage{dsfont}
\usepackage{mathtools}
\usepackage{url}
\usepackage{hyperref}
\hypersetup{colorlinks=true}
\usepackage{algorithm}
\usepackage{algorithmic}
\usepackage{adjustbox}
\usepackage{multirow}
\usepackage [operators,sets] {cryptocode}
\usepackage{bm}
\usepackage{enumitem}
\setlist{nosep} \setlist[1]{labelindent=\parindent}
\usepackage{qcircuit}

\usepackage{booktabs} 

\makeatletter
\AtBeginDocument{\def\doi#1{\url{https://doi.org/#1}}}
\makeatother

\usepackage{caption}
\usepackage{subcaption}

\floatstyle{ruled}

\newfloat{protocol}{htb!}{idf}
\floatname{protocol}{Protocol}

\newfloat{resource}{htb!}{idf}
\floatname{resource}{Resource}

\newfloat{simulator}{htb!}{idf}
\floatname{simulator}{Simulator}

\newfloat{problem}{htb!}{idf}
\floatname{problem}{Problem}

\usepackage{marginnote}

\usepackage{braket}
\usepackage[T1]{fontenc}
\usepackage[sc]{mathpazo}
\usepackage{comment}
\usepackage[colorinlistoftodos]{todonotes}

\newcommandx{\info}[2][1=]{\todo[linecolor=grey,backgroundcolor=grey!25,bordercolor=grey,#1]{#2}}
\newcommandx{\change}[2][1=]{\todo[linecolor=blue,backgroundcolor=blue!25,bordercolor=blue,#1]{#2}}
\newcommandx{\missing}[2][1=]{\todo[linecolor=red,backgroundcolor=red!25,bordercolor=red,#1]{#2}}

\newcommandx{\doubt}[2][1=]{\todo[linecolor=green,backgroundcolor=green!25,bordercolor=green,#1]{#2}}

\title{\huge Heuristic-free Verification-inspired \newline Quantum Benchmarking}

\author{
Johannes Frank \inst{1,4}, 
Elham Kashefi \inst{2,3,4}, 
Dominik Leichtle \inst{2},
Michael de Oliveira \inst{4,5,6}}
\institute{
Technical University of Munich, Germany  \and
School of Informatics, University of Edinburgh, 10 Crichton Street, EH8 9AB Edinburgh, United Kingdom  \and
National Quantum Computing Centre, Didcot, OX11 0QX, United Kingdom \and
Laboratoire d’Informatique de Paris 6, CNRS, Sorbonne Université, 4 Place Jussieu, 75005 Paris, France  \and 
International Iberian Nanotechnology Laboratory, Portugal\and INESC TEC, Braga, Portugal
}

\date{}

\begin{document}

\pagestyle{plain}
\large
\maketitle
\begin{abstract}
In this paper, we introduce a new approach to quantum benchmarking inspired by quantum verification motivating new paradigms of quantum benchmarking. Our proposed benchmark not only serves as a robust indicator of computational capability but also offers scalability, customizability, and universality. By providing formal statements regarding the quality of quantum devices while assuming device consistency, we eliminate the reliance on heuristics. We establish a deep connection between quantum verification and quantum benchmarking. For practical application, we present a concrete benchmarking protocol derived from a quantum verification protocol, and prove it to match our redefined standards for quantum benchmarking.

\vspace{1em} \noindent \textbf{Keywords:} Quantum Benchmarking, Quantum Verification.
\end{abstract}

\normalsize

\section{Introduction}\label{sec:intro}

Advances in quantum computation, mainly in hardware development, have taken significant steps toward the goal of building fault-tolerant quantum computers. With this constant improvement, it is of great interest to assess and compare the performance of various quantum devices. This benchmarking process plays a vital role in gauging the quality of ever-improving quantum hardware while simultaneously aiding end-users in determining the suitability of a specific quantum device for their intended applications. Currently, no universally accepted standard benchmarking technique exists. Moreover, so far, most considered quantum benchmarks have been largely heuristic and do not generate strong or reliable predictions about the computational performance of quantum devices. Recognizing the need for a more formal underpinning of predictive statements, recent efforts have emerged to strengthen the statements derived from heuristic quantum benchmarks, as discussed in \cite{hothem2023predictive}. From this standpoint, there is a clear desirability to transition benchmarking from its current heuristic nature to a more formal and predictable framework. 
Our aim is to make a meaningful contribution to this ongoing quest by integrating benchmarking with a previously distinct approach: quantum verification. 

As already pointed out in~\cite{Eisert_2020}, quantum benchmarking is naturally related to the field of quantum verification as both involve the certification of quantum computations.
As our main contribution, we establish a strong link between these two fields by presenting a generic way of compiling blind verification schemes to benchmarking protocols.
This compiler relies on a reduction from worst-case to average-case computations that blind verification schemes with strong security inherently exhibit.
We further prove that quantum benchmarking schemes obtained in this verification-inspired way achieve the new stronger paradigms.
We then apply this compiler to the robust quantum verification scheme of~\cite{Leichtle21} to obtain a concrete benchmarking protocol that inherits its favorable properties of scalability, efficiency and noise robustness. The employed verification techniques do not entail a quantum hardware overhead beyond the size of the target computation.
The new scheme does not rely on any heuristics or on any assumptions regarding the structure of noise and is inherently platform-agnostic.
It offers statements that can be seen as certificates of computational power rather than a heuristic performance indicator.

\subsection{Related work}

Within quantum benchmarking, several methodologies have been assessed. A significant portion of them does not fully meet the exhaustive criteria desired for a quantum benchmark. Drawing from our own work and the guidelines outlined in \cite{amico2023defining}, we conclude that an effective quantum benchmark should be unambiguous to clarify its derived metrics. It would ideally remain platform-agnostic, ensuring that its characterization of quantum capabilities is universal and not constricted to specific architectures or instructions. An essential facet is the ability of the benchmark to deduce optimal global properties, even when the data at hand is limited. That is, a benchmark should not vastly underestimate the capabilities of a device. Furthermore, it is crucial for a benchmark to offer predictive insights into a device's computational performance which are rigorously certified.

Building on these criteria, it is vital to analyse existing benchmarks with respect to their limitations. Some benchmarks are considerably specialized, focusing on certifying specific quantum processes or states rather than providing a global device evaluation \cite{mooney2021generation}. Prominent techniques like quantum volume and mirrored quantum volume shed light on a quantum device's deviations when executing a generic type of random unitary defined by specific width and depth in a circuit, as illustrated in \cite{cross2019validating,proctor2022measuring}. More recently, even more, fragmented approaches have been introduced to assess larger quantum devices with limited connectivity, resulting in certifications that are less robust \cite{MCKay23}. While they offer insights into how circuits of specific dimensions may be affected by noise, they do not provide the predictive power we deem desirable, especially regarding the fidelity of executing specific unitaries.  

Application-based benchmarks gauge quantum device performance across various quantum algorithms and processes. They pinpoint the performance thresholds and highlight the deficiencies of metrics like quantum volume when applied to practical quantum solutions. As \cite{lubinski2023applicationoriented} elucidates, errors influence circuit compilation of certain quantum routines differently than they do generic random unitary operations of similar dimensions. This distinction sometimes undermines predictions from quantum volume metrics. However, these benchmarks have their inadequacies, as they are not algorithm-oblivious, potentially leading to biased results for devices reacting differently to distinct algorithmic solutions.

The CLOPS benchmark, as cited in \cite{wack2021quality}, emphasizes the speed of quantum operations, shedding light on the time resources required for specific quantum tasks. 
This approach is motivated by classical benchmarking where the main focus of benchmarking lies on the speed of operations due to high confidence in the correct execution of operations. However, on quantum hardware fast execution alone cannot be used as an indicator for computational performance as all available devices struggle with noise.

The work of \cite{hothem2023predictive} underscores the potential of predictive benchmarks, utilizing both error models and neural networks. Notably, using quantum device data the error models excelled in determining realistic gate fidelities, sometimes diverging from manufacturer claims. Meanwhile, the neural network models predicted the success probabilities of specific unitaries, even ones not in the training data. However, both models have scalability questions unanswered and do not provide concrete guarantees on their predictions.
Lastly, verification protocols, as discussed in \cite{kapourniotis2022unifying,Leichtle21,eisert2020quantum}, offer a means to certify particular quantum algorithms or protocols with minimal overhead, even in the presence of noise. While yet to be adapted into general benchmarking with predictive attributes, they excel at offering a rigorous certification of a quantum algorithm's validity. These protocols harness the potential of blind delegation of computations, with embedded traps ensuring the algorithm's integrity. Recognizing deviations at these trap points offers insights into the overall fidelity of the process. We aim to integrate these techniques and protocols, crafting benchmarks that resonate with these attributes.

\subsection{New paradigms for quantum benchmarking}\label{new_para}

\vspace{0.2cm}
\noindent \textbf{Noise model agnostic.} To assess the performance of quantum devices, certain assumptions regarding the impact of noise on computations have been posited. These assumptions fix a distinct noise model that may not accurately reflect the actual noise effects present in the device. One illustration is the employment of gate fidelities, operating as an unital noise model. While they can offer predictions on the fidelity of a unitary or a circuit, they overlook certain quantum device issues like crosstalk, which typically amplifies the noise interference in computations. In this context, we aim to introduce a benchmark devoid of any noise model assumptions, accommodating all potential noise disturbances.

\vspace{0.2cm}
\noindent \textbf{Hardware agnostic.} The benchmarking of quantum devices should be independent of the hardware platform. Therefore, the benchmark should remain unaffected by any provider's specific operations/instructions and must be framed to ensure this. This independence is crucial, especially given the ongoing exploration of multiple hardware platforms to establish a sense of comparability across hardware platforms. It is worth noting that while these benchmarks and ours are crafted for the end user, benchmarks that are hardware-specific could prove valuable for hardware providers, assisting in the evolution of their quantum devices.

\vspace{0.2cm}
\noindent \textbf{Algorithm agnostic.} Benchmarks should not determine the capabilities of a machine solely based on a collection of quantum algorithms or routines. While some classical benchmarks are constructed this way \cite{lubinski2023applicationoriented}, our grasp of the quantum algorithmic landscape is still evolving, and recent discoveries of algorithms exhibiting exponential quantum advantages \cite{yamakawa2022verifiable,babbush2023exponential} indicate the potential for a spectrum of quantum routines that offer computational benefits. Hence, it is premature to design machines optimized solely for the current known algorithms, potentially overlooking emergent solutions. A benchmark should aim at providing strong statements about the performance of the device that should be useful as indicator across a multitude of possible applications. This underscores an advantage of our benchmark: it can determine the potential of new algorithmic approaches based solely on their circuit size/scaling, independent of specific instructions or architecture.

\vspace{0.2cm}
\noindent \textbf{Predictive $\&$ Certifiable.}  Quantum benchmarks should not only rank quantum devices based on their efficacy in executing quantum computations but also offer insights into the performance of other quantum computations and routines one might wish to deploy on those devices. Essentially, they should globally profile the machine, equipping the end user with insights into the device's performance across diverse and user-specified applications. Beyond the given predictive capacity, there should be a level of assurance, or to put it more directly, a guarantee of accuracy. This ensures that the predictions align with a recognized benchmark. Our method fulfills both these criteria, differentiating it from previous solutions.

\vspace{0.2cm}
\noindent \textbf{Explicit, Practical $\&$ Scalable.}
Lastly, a benchmark should be detailed, providing a straightforward set of instructions to execute without being overly complex. 
Also, it should not incur significant computational overheads that lead to an underestimation in the assessment of quantum devices.
Finally, the overall computational resources demanded by the benchmark should grow at most polynomially, guaranteeing its feasibility and scalability beyond small-sized quantum devices.

\subsection{Organization}
In section \ref{sec:prelim} we introduce the technical definitions that we will work with throughout the paper. Notably, we formalize the behavior of quantum devices, which is needed to prove heuristic-free statements, and define quantum verification protocols in an abstract way. In section \ref{sec:formal_treatment_of_benchmarking} we introduce a formal definition of a certifiably implementable class of computations (CICC) benchmark. We elaborate on which properties each of them should have to investigate the computability of a class of computations on a given quantum device. In addition, we prove that from any verification scheme one can efficiently compile a CICC benchmark. In section \ref{sec:concrete_efficient_benchmarking_protocol} we instantiate a concrete efficient CICC benchmark utilizing the robust VBQC protocol from \cite{kapourniotis2022unifying}.

\section{Preliminaries}\label{sec:prelim}

In this section, we aim to present a rigorous examination of the primary subject of our analysis. Initially, we will describe quantum devices, outlining only the essential requirements to ensure the hardware-agnostic proprieties. Following this, we will define quantum verification protocols, which will serve as the tools to assess the performance of quantum devices.

\subsection{Quantum devices}
\label{subsec:quantum_devices}
In Quantum Benchmarking we want to assess the performance of a quantum device carrying out tasks that involve quantum information processing. For the end-user it is favorable not to make any assumptions on the quantum device and treat it as a black box to get unbiased performance indicators. 

The benchmark that we propose justifies its statements about quantum devices formally which requires a formal description of the behavior of physical quantum devices. We introduce an interactive quantum algorithm (Definition \ref{def:interactive_quantum_algorithm}) that describes the action of an a general quantum device partaking in an interactive protocol.
The formalism we choose in this work is not unique but it covers the any behavior a quantum device can exhibit in a protocol type setting. We only consider the interfaces and treat the devices themselves as a black box because we do not want to make assumptions about the the inner workings of a device to gauge performance in carrying out a desired task. 
An interactive quantum algorithm is designed to be part of a most general type of quantum protocols: In between the rounds of communication the parties/devices can act on the received quantum states with quantum channels and use an arbitrary quantum and classical memory to store information and entangle states from round to round. Classical information is also encoded in quantum states in this picture.
\begin{definition}[Interactive Quantum Algorithm]
\label{def:interactive_quantum_algorithm}
Let $\mathcal{I}_i$, $\mathcal{O}_i$, and $\mathcal{D}i$ denote the input quantum states, output quantum states, and internal memory states of the quantum device in each round, respectively. Consider a sequence of $n$ quantum channels, $\{\mathcal{E}^{i} : \mathcal{S}(\mathcal{I}_i \otimes \mathcal{D}_{i-1}) \longrightarrow \mathcal{S}(\mathcal{O}_i \otimes \mathcal{D}_{i})\}_{i=1}^n$ represents the space of all density matrices over the Hilbert space $\mathcal{H}$. We define an algorithm that iteratively executes the following steps for each $i$ from 1 to $n$:\newline

1. \textbf{Receive Input}: $\rho_I^{i}$\newline

2. \textbf{Return Output}: $\rho_O^{i}=\mathrm{Tr}_{\mathcal{D}_i}[\mathcal{E}^{i}(\rho_I^{i} \otimes \sigma^{i-1})]$\newline

3. \textbf{Update Internal Memory}: $\sigma^{i}=\mathrm{Tr}_{\mathcal{O}_i}[\mathcal{E}^{i}(\rho_I^{i} \otimes \sigma^{i-1})] $\newline

\noindent an interactive quantum algorithm $\mathbf{Q}$. 

\end{definition}

The motivation for this very technical definition and somewhat trivial summary of interactive quantum algorithms is that we will need formal objects that can model the entire behavior of a physical quantum device to make formal assumptions and statements about their performance.
 One fundamental assumption that underlies our benchmark is that the interactive quantum algorithm $\mathbf{Q}$ that a quantum device implements stays the same over all interactions with the device. The formalism still allows for randomness in the behavior (quantum channels constituting $\mathbf{Q}$) but the underlying quantum process must be consistent. This might seem like a very strong assumption but in fact it is necessary for any kind of benchmark: if the benchmarked device is changing after the benchmark, we have no information about the new behavior. 
 
 Note that we will also refer to the quantum device itself as $\mathbf{Q}$. The difference of the quantum device to the interactive quantum algorithm is merely that a device repeats the same interactive quantum algorithm $\mathbf{Q}$ multiple times, just like a physical device would. We expect a device $\mathbf{Q}$ to act in the same way everytime we interact with it. 
 It is also worth noting that not only the behavior of the quantum device that is to be benchmarked can be formalized like in definition \ref{def:interactive_quantum_algorithm} but actually the entire benchmarking protocol we cover in this work will follow this formalism.

However, we will not use this tedious notation to define the interactive algorithms that come up in this work but we will stick to a more intuitive, graphical depiction instead, which can be seen in figure \ref{fig:verification_visualization}.

When interacting two quantum algorithms we use the notation $\langle \mathbf{A},\mathbf{B}\rangle$. Whenever we use this notation it is implicitly assumed that the interfaces of $\mathbf{A}$ and $\mathbf{B}$ match. The resulting structure will again be an interactive quantum algorithm if it has an open input/output interface on one or both sides. In favor of readability this will sometimes be neglected while at other points we will use the notation of an inward facing arrow for inputs and an outward facing arrow for outputs like so: 
\begin{equation}
    \mathrm{output}\leftarrow\langle \mathbf{A},\mathbf{B}\rangle \leftarrow \mathrm{input}
\end{equation}

It is also possible, if the communication rounds of $\mathbf{A}$ fit $m$ times the one of $\mathbf{B}$, to interact multiple versions of the same algorithm which we denote as $\langle \mathbf{A}, \mathbf{B}^m\rangle$.

\subsection{Quantum Verification Protocols}
\label{subsec:quantum_verification}
Quantum verification protocols are designed for delegating quantum computations in a secure and certifiable way to a server or a device. To make them accessible for our benchmark we introduce verification in an interactive quantum algorithm picture.
We define the client side of a verification protocol as interactive quantum algorithm $\mathbf{V}$ which delegates a computation from a class of computations: $C \in \mathfrak{C}$ 
via interacting with a quantum device $\mathbf{Q}$.  
With a computation $C$ we mean a classical description of a quantum computation, in most general terms: a CPTP map.

\begin{definition}[Verification Algorithm]
\label{def:verification_algorithm}
    We call an interactive quantum algorithm $\mathbf{V}$ that:
    \begin{enumerate}
    
        \item Takes input $C$, the classical information describing a quantum computation in class of computations $\mathfrak{C}$
        \item Interacts with a quantum device $\mathbf{Q}$
        \item Outputs the result of the computation and a flag indicating whether to accept or reject the result
     \end{enumerate}
     
     At the same time it fulfills:
     
        \begin{enumerate}[label=\alph*)]
            \item 
            \begin{equation}
            \label{eq:verification_property}
                \exists \delta > 0 : \forall C \in \mathfrak{C}, \forall \mathbf{Q}: \Pr[\mathrm{acc}\land \mathrm{wrong}\mid (\mathrm{res},\mathrm{flag})\leftarrow\langle \mathbf{V},\mathbf{Q}\rangle \leftarrow C] \leq \delta
            \end{equation}
            \item 
            \begin{equation}
            \label{eq:success_of_verification_protocol}
                \exists \mathbf{P}: \forall C \in \mathfrak{C}: \Pr[\mathrm{acc}\mid (\mathrm{res},\mathrm{flag})\leftarrow\langle \mathbf{V},\mathbf{P}\rangle \leftarrow C] = 1
            \end{equation}
        \end{enumerate}
  
    a verification algorithm $\mathbf{V}$
\end{definition}

\noindent \textbf{Condition \textit{a}}. Guarantees that there is an upper bound $\delta > 0$ on the probability of erroneously accepting a wrong result. This upper bound should preferably approach zero in some parameter for example when scaling the size of the verification scheme.

\vspace{0.2cm}
\noindent \textbf{Condition \textit{b}}.  Makes sure that a verification scheme is accompanied by a prover algorithm $\mathbf{P}$, representing an ideal quantum device flawlessly implementing the server side of the scheme. Also, it excludes schemes that reject always and thus trivially fulfill \ref{eq:verification_property} while not being useful at all.

\begin{figure}[h]
    \centering
    \hfill
    \begin{subfigure}[t]{0.4\textwidth}
      \centering
      \includegraphics[width=\textwidth]{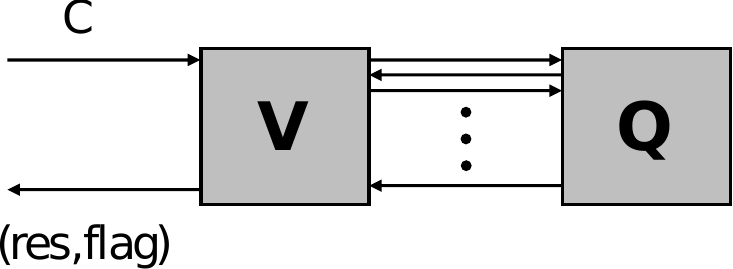}
      \caption{Quantum verification as interactive quantum algorithm. This first input is a classical description of a quantum computation $C \in \mathfrak{C}$. Then it interacts with a quantum device $\mathbf{Q}$ to delegate the computation. As a last output it provides the result of the computation and a flag indicating accept or reject.}
    \end{subfigure}
    \hfill
    \begin{subfigure}[t]{0.4\textwidth}
      \centering
      \includegraphics[width=\textwidth]{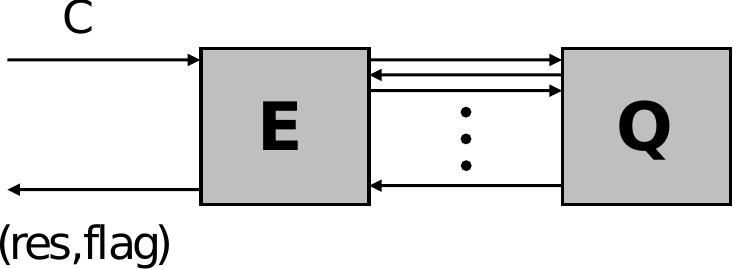}
      \caption{Extractor Algorithm $\mathbf{E}$ delegating computation $C$ to quantum device  $\mathbf{Q}$ yielding result and flag: $(\mathrm{res},\mathrm{flag})\leftarrow \langle \mathbf{E},\mathbf{Q}\rangle\leftarrow C$. The first input computation $C\in \mathfrak{C}$ is meant to come from the client and the result and flag are returned also to the client.}
    \end{subfigure}
    \hfill
  \caption{Quantum verification and extraction as interactive algorithms.}
  \label{fig:verification_visualization}
\end{figure}

Provided the formal treatment of the verification algorithms that will be employed we do point to figure \ref{fig:verification_visualization} for a graphical representation of the structure of these processes. 

\section{A formal treatment of benchmarking}
\label{sec:formal_treatment_of_benchmarking}
On a more abstract level, benchmarking can be described as the characterization of quantum processes. The most comprehensive form of this characterization is quantum process tomography, which, unfortunately, is not feasible due to poor scaling behavior. However, benchmarking can be viewed as an attempt to ascertain the quantum information processing capabilities of a quantum device using fewer resources than tomography requires. To achieve this, a practical benchmark is divided into two stages. In the first stage, the client runs an efficient benchmarking algorithm, $\mathbf{B}$, interacting with the quantum device, $\mathbf{Q}$, that is undergoing benchmarking. The second stage involves delegating actual computations to the quantum device via an extractor algorithm, $\mathbf{E}$. The extractor serves as a set of instructions, specifically the quantum channels the client uses to convey their quantum computations to the device. The insights gained by the client in the first stage provide a lower bound on the success probability of executing computations belonging to class $\mathfrak{C}$. Any non-zero lower bound on this success probability indicates that computations of class $\mathfrak{C}$ can be made computable through repetition. 

Expanding upon this initial overview, we will refine the precise definitions, properties, and their relationships to verification, which will be segmented into the following subsections.
\vspace{0.2cm}

\begin{itemize}
    \item In section \ref{subsec:definitions} we make the formal foundation for our benchmark. We start by defining the form of extractor algorithm which makes concrete how the client will implement computations on a device. Secondly, we introduce the form of our benchmark in the picture of interactive quantum algorithms, without fixing any properties yet.
    \item We go on in section \ref{subsec:properties_of_a_cicc_benchmark} to define properties of a before defined benchmark that are desirable. The properties are chosen such that it will be easy to prove them in the following section. The important part is what follows from fulfilling all of them which will also be explained. 
    \item Finally, in section \ref{subsec:cicc_benchmark_from_verification} we introduce a scheme of utilizing multiple verification protocols that qualifies as a CICC benchmark in the sense that we defined. We go on to show that it fulfills all the properties that we want from the benchmark to be useful.
\end{itemize}

\subsection{Definitions}
\label{subsec:definitions}

The goal of any benchmark is to make statements about the performance of quantum devices on tasks that require quantum information processing. The exact statements that we will make concern what we call the computability of computational classes. Additionally, we aim to remove any heuristics from our benchmark. We achieve this by introducing the extractor algorithm to anchor the computability statements: If a device $\mathbf{Q}$ is certified via our benchmark for a class of computations $\mathfrak{C}$, then there exists an interactive quantum algorithm (the extractor $\mathbf{E}$) which the client can use to delegate any computation in $\mathfrak{C}$ to the device and is guaranteed formally to get the correct result in at most a small number of tries with no heuristic arguments involved. For a more intuitive understanding of what the extractor algorithm does, we include a graphical depiction of it in figure \ref{fig:verification_visualization} b). Furthermore, we define the interactive quantum algorithm that the client ideally expects from the device in the delegation protocol as the prover algorithm $\mathbf{P}$. Thus, we assume $\langle \mathbf{E},\mathbf{P}\rangle$ perfectly implements any computation in $\mathfrak{C}$. We can now define our proposed benchmark.

\begin{definition}[Certified Implementable Class of Computations Benchmark]
\label{def:cisc_benchmark}
Let $\mathfrak{C}$ be a class of computations. We call the tuple of interactive quantum algorithms $(\mathbf{B}, \mathbf{E}, \mathbf{P})$ where
\begin{itemize}
        \item $\mathbf{B}$, referred to as the benchmarking algorithm, engages in interactions with a quantum device $\mathbf{Q}$ for a total of $m$ times, ultimately producing an output $z \in [0, 1] \leftarrow \langle \mathbf{B}, \mathbf{Q}^m\rangle$.
        \item $\mathbf{P}$, known as the prover algorithm, outlines the expected behavior of a quantum device within the delegation scheme of $\mathbf{E}$.   
        \item $\mathbf{E}$, functioning as the extractor algorithm, takes input $C \in \mathfrak{C}$, engages in interactions with a quantum device, and ultimately produces the result along with a flag indicating acceptance: $(\mathrm{res}, \mathrm{flag}) \leftarrow \langle\mathbf{E},\mathbf{P}\rangle$. It consistently accepts and is accurate for all $C \in \mathfrak{C}$.
\end{itemize}
a Certified Implementable Class of Computations Benchmark.
\end{definition}

\begin{figure}
  \centering
  \includegraphics[width=0.6\textwidth]{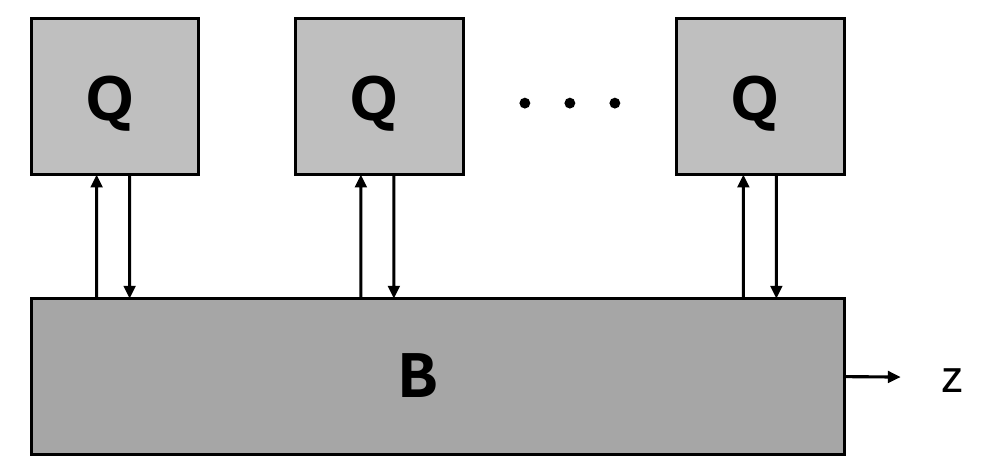}
  \caption{Benchmarking Algorithm $\mathbf{B}$ interacting with a quantum device $\mathbf{Q}$ $m$ times outputting information $z\in [0,1]$ that is based on $\mathbf{Q}$.}
  \label{fig:benchmark_visualization}
\end{figure}

We implicitly assume that all three algorithms that constitute the benchmark match on their interfaces. 
The structure of the benchmarking algorithm is depicted in figure \ref{fig:benchmark_visualization}.

The final goal of our benchmark will be to output a $z \in [0,1]$ that is a lower bound on the success probability of running any computation $C \in \mathfrak{C}$ on $\mathbf{Q}$ via the extractor $\mathbf{E}$. However, at this point it is not clear yet how this property will be achieved. As of now, we merely created the abstract form of a CICC benchmark in a protocol picture. In the next section (\ref{subsec:properties_of_a_cicc_benchmark}) we will define properties that a CICC benchmark should fulfill. It becomes useful for practical applications only if all the properties are really fulfilled. For now we want to give a few more insights on the concepts that are already introduced.

\vspace{0.2cm}
\noindent 
\textbf{Classes of Computations}
The class of computations $\mathfrak{C}$ is always linked to the choice of the extractor algorithm. For example, any delegation scheme (extractor) of finite size necessarily restricts the size of computations. In this work we also restrict everything to classical input / classical output quantum computations.
Note that we only handle the classical description $C$ of these computations in the communication to the extractor and we assume this description to be efficient.

\vspace{0.2cm}

When comparing devices with a CICC benchmark, the idea is to check which device can successfully implement a larger set of computations $\mathfrak{C}$ or, as the sets might not be inclusive, which device can implement a set that includes a desired subset of computations $\mathfrak{c}$. 

\vspace{0.2cm}
\noindent \textbf{Benchmarking Stage}.
For the CICC benchmark it is important that the quantum device $\mathbf{Q}$ stays the same through out the interaction with $\mathbf{B}$ as well as extending to the computational stage when the client performs computations via $\langle \mathbf{E},\mathbf{Q}\rangle$. Every interaction with the device must implement the same interactive quantum algorithm $\mathbf{Q}$ on the device side. In particular, a device has no historic data to adapt its strategy. This is vital for the integrity of the benchmark.

\vspace{0.2cm}
\noindent \textbf{Extractor Algorithm}.
The choice of the extractor is crucial. The more resources the extractor uses, the less powerful the statement about the quantum device is. That is because the more quantum resources the extractor uses to delegate a computation the less quantum resources need to be employed on the device side to implement computations. Thus the overhead of the extractor is important. Also, the resources which a client has access to are determining which extractor to use in a benchmark.

\subsection{Properties of a CICC Benchmark}
\label{subsec:properties_of_a_cicc_benchmark}
In this section, we introduce the essential properties that a CICC benchmark must fulfill and explain why these properties lead to a useful benchmark. The first property we discuss is correctness, which ensures that an optimal device receives an optimal score with high probability. This is a fundamental expectation of any effective benchmark, as it stands to reason that a perfect device should score perfectly.

\begin{definition}[Correctness]
\label{def:correctness}
    We say Benchmark $(\mathbf{B}, \mathbf{E},\mathbf{P})$ is $\alpha$-correct for a shift of $\beta$ if 
\begin{equation}
    \label{eq:correctness}
    \Pr[z \geq 1-\beta \mid z \leftarrow \langle \mathbf{B}, \mathbf{P^m} \rangle] \geq 1- \alpha
\end{equation}
\end{definition}

The reason why we introduce the $\beta$-shift is to add some kind of robustness. For a process that inherently involves randomness we do not want to include a hard condition such as $z=1$. It also suffices to ensure there is a high probability $(\geq 1-\alpha)$  to score close to optimal $(\geq 1-\beta)$ with the optimal device $\mathbf{P}$. 

Recall from the definition of the CICC benchmark that the ideal device implementing the prover algorithm $\mathbf{P}$ is simultaneously the perfect counterpart to the extractor algorithm $\mathbf{E}$, which, by definition, has a success probability of exactly 1 and should, therefore, score very high on the benchmark. Building on this, the next two properties introduce what we call the extractability of the benchmark. This property aims to encapsulate the client's ability to obtain correct results, which are also flagged as "accept", for the computations they implement ensuring that the benchmark not only measures performance but also the practical usability of the device in delivering reliable outcomes. 

\begin{definition}[Correctly-Flagged Extractability]
\label{def:correctly_flagged_extractability}
We say a benchmark $(\mathbf{B}, \mathbf{E},\mathbf{P})$ is $\gamma$-correctly-flagged extractable if for all $\mathbf{Q}$ of the form of $\mathbf{P}$ it holds that:
\begin{equation}
       \forall C \in \mathfrak{C}: 
        \Pr[\Pr[\mathrm{flag} = \mathrm{acc}\mid (\mathrm{res},\mathrm{flag})\leftarrow \langle\mathbf{E}, \mathbf{Q} \rangle ] \leq z \mid z \leftarrow \langle\mathbf{B}, \mathbf{Q}^m \rangle] \leq \gamma
\end{equation}

\end{definition}

This property of a CICC benchmark refers to the computation stage where the client implements a computation $C\in \mathfrak{C}$ via the extractor $\mathbf{E}$ on the quantum device $\mathbf{Q}$. It makes sure that is unlikely (small $\gamma$) that the acceptance rate is lower than the value $z$ that the benchmark is suggesting. Also, note that this will be the only property that actually links the benchmarking algorithm $\mathbf{B}$ to the extractor algorithm $\mathbf{E}$. 

We are now only missing a property that we call verifiable extractability:

\begin{definition}[Verifiable Extractability]
\label{def:verifiable_extractability}
We call benchmark $(\mathbf{B}, \mathbf{E},\mathbf{P})$ $\delta$-verifiably extractable if for all $\mathbf{Q}$ of the form of $\mathbf{P}$ it holds that:
\begin{equation}
        \forall C \in \mathfrak{C}: 
        \Pr[\mathrm{res} =  \mathrm{wrong} \land  \mathrm{flag} = \mathrm{acc} \mid (\mathrm{res},\mathrm{flag})\leftarrow  \langle\mathbf{\mathbf{E}}, \mathbf{Q} \rangle
        ]  \leq \delta
    \end{equation}

\end{definition}
We use the notation "$\mathrm{res} = \mathrm{wrong}$" to denote the stochastic event where the result returned by the interaction between the extractor and the device, given an input $C \in \mathfrak{C}$, is incorrect. Since we are dealing with classical input/classical output quantum computations, the distinction between "correct" and "wrong" results is clearly defined. This property pertains exclusively to the extraction algorithm, ensuring that the likelihood of obtaining an incorrect result that is mistakenly marked as correct is low (denoted by a small $\delta$).\newline

If a CICC benchmark fulfills properties \ref{def:correctly_flagged_extractability} and \ref{def:verifiable_extractability} then the retrieved value of $z$ can be used as a lower bound on the probability of getting a (flagged) correct result on a computation. To see this we follow the subsequent argument \footnote{Note that we have introduced an abbreviated notation where "acc" refers to the probability of success for implementing a computation via the extractor.}.

Correctly-flagged extractability \ref{def:correctly_flagged_extractability} gives us directly:
\begin{equation}
\label{eq:correctly_flagged_extractability}
    \Pr[\Pr[\mathrm{acc}]> z ] > 1-\gamma
\end{equation}
At the same time verifiable extractability \ref{def:verifiable_extractability} gives us:
\begin{equation}
\label{eq:verifiable_extractability}
    \Pr[\mathrm{wrong} \land \mathrm{acc}] \leq \delta
\end{equation}

\noindent Now we use:
\begin{equation}
\label{eq:logical_and_summation}
    \Pr[\mathrm{acc}] = \Pr[\mathrm{correct} \land \mathrm{acc}] + \Pr[\mathrm{wrong} \land \mathrm{acc}] 
\end{equation}

\noindent Combining equations \ref{eq:correctly_flagged_extractability} - \ref{eq:logical_and_summation} we find:
\begin{equation}
    \Pr[\Pr[\mathrm{correct} \land \mathrm{acc}] > z-\delta] > 1-\gamma
\end{equation}

This means that $z$, retrieved from the benchmark, shifted by $\delta$, serves as a lower bound for the fraction of accepted correct results with probability of $1 - \gamma$. If a CICC benchmark meets all these criteria and returns a non-zero success probability $z$, then by repeating the computation with $\mathbf{E}$ multiple times, the likelihood of not obtaining a single correct result decreases exponentially with the number of runs, provided $z - \delta > 0$. The convergence speed depends on $z$.

Even though the original output of the benchmark is the lower-bound $z$, it is more interesting to note that as long as the success probability is non-zero then, with a small overhead in time (and not in the size of the quantum device), the device can certainly carry out every computation from the class $\mathfrak{C}$. Thus, one can hand out a certificate of computability for a device that was benchmarked in this way.

\vspace{0.2cm} 
\noindent \textbf{Remark.} We want to stress that a specific benchmark that comes with values of $\delta$ and $\gamma$ is only useful if there is a systematic way how to make these values approach zero exponentially fast. The benchmark that we suggest will achieve this via a repetition scheme where both $\delta$ and $\gamma$ will be negligible in the number of rounds.

\subsection{CICC Benchmark from Verification}
\label{subsec:cicc_benchmark_from_verification}

In this subsection, we delve into the process of constructing a CICC benchmark based on an arbitrary verification scheme, $\mathbf{V}$, as introduced in section \ref{subsec:quantum_verification}, which is pertinent to classical input classical output quantum computations. The core concept is fairly simple: we employ a verification scheme repeatedly to estimate its success rate. Following this, we delineate the CICC benchmark formally as $(\mathbf{B}_{\mathbf{V}}, \mathbf{V},\mathbf{P} )$, ensuring it fulfills the critical properties \ref{def:correctness} - \ref{def:verifiable_extractability},  and explain the constituents subsequently.

\begin{figure}
  \centering
  \includegraphics[width=0.6\textwidth]{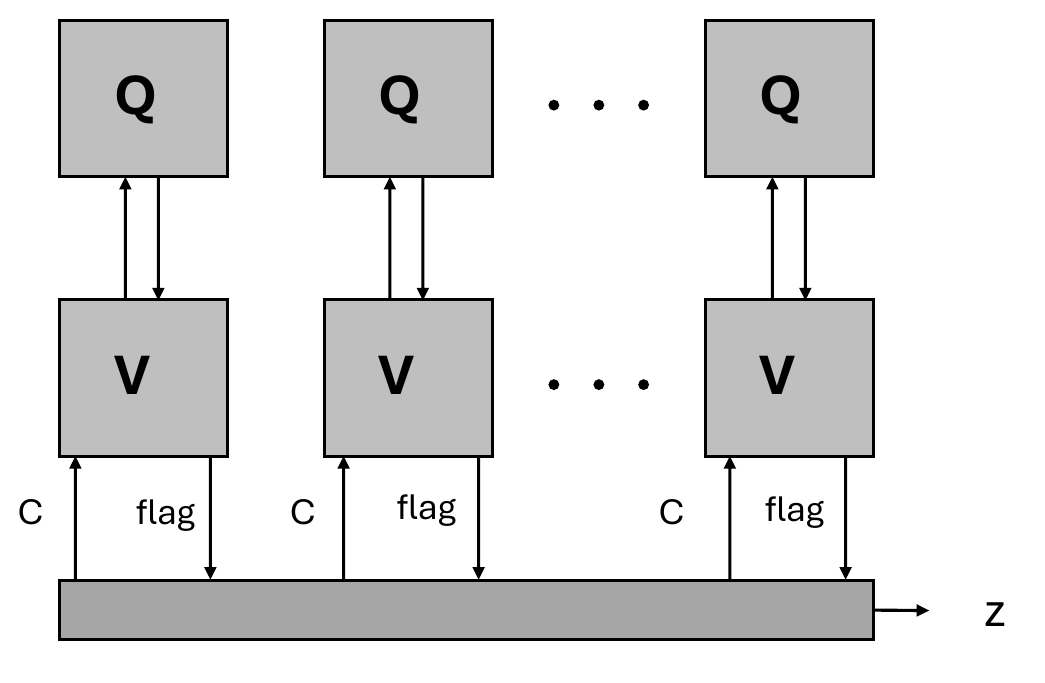}
  \caption{Verification-inspired CICC benchmarking algorithms $\mathbf{B}_{\mathbf{V}}$ utilizing $m$ rounds of the verification algorithm $\mathbf{V}$}
  \label{fig:verification_benchmark_visualization}
\end{figure}

\vspace{0.2cm}
\noindent \textbf{Benchmarking algorithm $\mathbf{B}_{\mathbf{V}}$}. The benchmark repeats verification protocol  $\mathbf{V}$ $m$ times on quantum device $\mathbf{Q}$. Then, it returns $z\in [0,1] = \frac{\#acc}{m}-\beta$ the fraction of accepted results, shifted by $\beta > 0$ so that $z$ serves as a lower bound on the success probability estimate of the verification protocol $\mathbf{V}$. We provide a schematic of the verification based benchmarking algorithm in\ref{fig:verification_benchmark_visualization}.

\vspace{0.2cm}
\noindent \textbf{Extractor algorithm $\mathbf{V}$.} The extractor algorithm which is needed to define the CICC benchmark is simply the verification protocol: $\mathbf{E} = \mathbf{V}$.

\vspace{0.2cm}
\noindent \textbf{Prover algorithm $\mathbf{P}$.} The prover algorithm $\mathbf{P}$ is the same that comes with the verification protocol, which was explained in section \ref{subsec:quantum_verification}. 
\vspace{0.2cm}

We now go on to prove that the benchmark defined by $(\mathbf{B}_{\mathbf{V}},\mathbf{V} ,\mathbf{P})$ fulfills properties \ref{def:correctness} - \ref{def:verifiable_extractability}.

\begin{proof}[Proof of Correctness]
\label{proof:correctness}
    Per definition, a verification protocol has an ideal counterpart algorithm that it can interact with perfectly i.e. $\mathbf{P}: (res,flag) \leftarrow \langle \mathbf{V}, \mathbf{P}\rangle$ gives the correct result and accepts at the same time deterministically. Therefore we trivially get that $\#acc = m$ and thus $z=1-\beta$ will be returned deterministically. Here, we made the choice of $\beta $ in the definition of the CICC benchmark deliberately such that it matches the one used in the correctness property. We get: 
    \begin{equation}
        \Pr[z\geq 1 - \beta \mid z \leftarrow \langle \mathbf{B}_{\mathbf{V}}, \mathbf{P}^m \rangle ] \geq \Pr[z=1 -\beta \mid z \leftarrow \langle \mathbf{B}_{\mathbf{V}}, \mathbf{P}^m \rangle ] =1 \geq 1- \alpha 
    \end{equation}
    This makes the benchmark $\alpha$-correct with shift $\beta$ for any $\alpha \geq 0$.
\end{proof}

\begin{proof}[Proof of Correctly-Flagged Extractability]

During the Benchmark we execute $\langle \mathbf{V}, \mathbf{Q} \rangle$ m times. Both interactive quantum algorithms  $\mathbf{V}$ and $ \mathbf{Q}$ stay the same that means that the probability which underlies the sampling of $flag \leftarrow \langle \mathbf{V}, \mathbf{Q} \rangle $ stays the same everytime. Also it is determined fully by $\mathbf{V}$ and $\mathbf{Q}$ independent of the computation $C$ which is chosen in the beginning of the benchmark due to the blindness of the verification protocol. This means that we effectively have m rounds of sampling the same, independent binary probability distribution of accept/reject. The Hoeffdings inequality for IID binary binomial distribution is explained in Appendix \ref{sec:hoeffding}. We identify accept with the outcome 1 and reject with the outcome 0. Then we are exactly in the case that is discussed in the appendix. We can just take the inequality that results for m times repeating a biased coinflip with probability $p$ of obtaining 1 and threshold $\beta$: $0 < \beta < \frac{\# acc}{m}$ and $z = \frac{\# acc}{m} - \beta$: 
\begin{equation}
\label{eq:hoeffding_p_z_t}
    \Pr[p\leq z]\leq \exp(-2m\beta^2)
\end{equation}
 Where random variable $z \in [0,1]$ is the averaged value after sampling m times the same coin flip shifted by small $\beta$ with unknown probability $p$.
 
 Note that in the case of our benchmark $p$ the probability of getting result 1 is assigned to the probability of getting an accept. Thus we will replace $p$ in equation \ref{eq:hoeffding_p_z_t} with  $\Pr[\mathrm{flag} = \mathrm{acc} \mid \mathrm{flag} \leftarrow \langle \mathbf{V}, \mathbf{Q}\rangle]$ . At the same time we identify the threshold parameter $\beta$ as the one from correctness. This gives us:
 \begin{equation}
 \label{eq:hoeffding_applied_to_extractability}
     \Pr[\Pr[\mathrm{flag}  =  \mathrm{acc} \mid (\mathrm{res}, \mathrm{flag}) \leftarrow \langle  \mathbf{V}, \mathbf{Q}\rangle]\leq z \mid z \leftarrow \langle   \mathbf{B}_{\mathbf{V}}, \mathbf{Q}^m\rangle] \leq \exp(-2m\beta^2)
 \end{equation}
Which is nothing else than $\gamma$-correctly-flagged extracability for $\gamma = \exp(-2m\beta^2)$

\end{proof}

The effect of $\beta$ is the following: If we want to find out the highest lower-bound on the success probability we will decrease $\beta$ but to achieve the same confidence we will then need a squared increase in rounds which comes from the exponent in \ref{eq:hoeffding_applied_to_extractability}. Usually $\beta$ does not need to go to zero though because a finite $\beta$ will suffice to obtain a reasonable lower-bound.

\begin{proof}[Proof of Verifiable Extractability]
\label{proof:verifiable_extractability}
The $\delta$-verifiable extracability property comes directly from the $\delta$-verifiability of the verification protocol that is used as explained in \ref{subsec:quantum_verification}.

\end{proof}

\subsection{Optimized Benchmarking Protocol from Robust VBQC}
\label{sec:concrete_efficient_benchmarking_protocol}

In this section, we adapt the generic CICC benchmark (section \ref{subsec:cicc_benchmark_from_verification}) using the robust VBQC protocol from \cite{kapourniotis2022unifying}. Due to the round-based structure of the robust VBQC protocol, we can reduce the time overhead during the benchmarking stage to get a slightly more efficient version of the generic CICC benchmark.

Notice that to avoid confusion, we stick to the following notions. When talking about "rounds" we refer to the sub-graph rounds in the robust VBQC MBQC pattern. The number $n$ will always refer to the number of rounds that constitute the verification scheme. Also, when referring to "repetitions" we mean the number of times a benchmarking algorithm is repeating a verification scheme or interacting with a quantum device $m$ times.

\subsubsection{Recap: robust VBQC protocol}
\label{subsec:recap_robust_vbqc_protocol}

The robust VBQC protocol was introduced in \cite{kapourniotis2022unifying}. It is written in the Measurement Based Quantum Computing framework \cite{PhysRevLett.86.5188} where computations are mapped to feed forward measurement sequences on large graph states.
 It is a provably secure (in the Abstract Cryptography framework) protocol to delegate quantum computations from a client that can only prepare and send single qubit states to a server that can prepare large scale entangled resource states (graph states) and use the clients input qubits to calculate the desired computation via feed forward measurements. 
The robust VBQC protocol uses the universal blind computation protocol from \cite{broadbent2009universal} to delegate computations in a completely blind way, making it impossible for the server to distinguish between different interactions as long as the underlying graph state remains the same.

\paragraph{The Protocol}
The Robust VBQC protocol has a round based structure. For $n$ rounds the client implements a blind MBQC pattern where it delegates its target computation to the server on the same graph state (subgraph $G$). A fraction of the rounds are randomly chosen by the client to feature a trap computation instead. Those are computations with a deterministic outcome known to the client. If the server's result deviates from the expected one the server's malicious or faulty behavior is revealed. As all the rounds are indistinguishable for the server, deviations must affect trap and computational rounds equally. The indistinguishability leads to an effective error model on subgraph $G$ that is sampled $n$ times that is, the error probability distribution is identical and independent in each round.
After running $n$ rounds (computations and traps) the client counts the number of total activated traps. If the trap activation rate is below threshold $c_t\cdot n$ the client accepts the result, performing a majority vote over all computational rounds.
The trap computations are chosen in such a way that the probability of the joint event of the trap activation rate being below the threshold (i.e. client accepting result) while the result of the majority vote is actually not the correct result of the desired computation is negligible in $n$:
\begin{equation}
    \Pr[\mathrm{accepts} \land \mathrm{wrong} ] \propto \exp(-n^2)
\end{equation}
    
Thus, a small overhead in time, not in the size of the quantum system (in the form of subgraph $G$), yields exponential security in the correctness of an accepted computation. Thus, the protocol can verifiably delegate quantum computations to the server without hardware overheads.

\subsubsection{Optimizing the Generic CICC Benchmark for Robust VBQC}\label{CICCinstance}

We propose here a concrete efficient benchmarking protocol that utilizes the round-based robust VBQC protocol from \cite{kapourniotis2022unifying}. It slightly differs from the generic case that we investigated in section \ref{subsec:cicc_benchmark_from_verification} in favor of only using trap computations.

We choose to instantiate our CICC benchmark from \ref{subsec:cicc_benchmark_from_verification} with the robust VBQC protocol because it is efficient. It only uses a small overhead in time and not in the size of the quantum system. Also, the round-based structure allows for a slight optimization that improves the confidence level. We could instead simply use the generic CICC benchmark from section \ref{subsec:cicc_benchmark_from_verification} to construct from robust VBQC protocol $ \mathbf{V}_n$ a benchmark $(\mathbf{B}_{\mathbf{V}_n}, \mathbf{V}_n,\mathbf{P})$. However, this only achieves a $\gamma$-correctly-flagged extracrability with $\gamma = \exp(-2m\beta^2)$, where the $n$-round verification protocol $\mathbf{V}_n$ is repeated $m$ times. We will show that there is an optimized version of the benchmark which achieves $\gamma = \exp(-2 mn\beta_t^2)$ decreasing the number of times $m$ we need to run the verification protocol to get the same confidence in the result of the benchmark.\newline

\noindent \textbf{Optimized CICC Benchmark from robust VBQC} 

We instantiate a CICC benchmark $(\mathbf{B}_{\mathbf{V}_n}, \mathbf{V}_n,\mathbf{P})$ and slightly alter the benchmarking algorithm $\mathbf{B}_{\mathbf{V}_n}$:
\begin{itemize}

        \item For all $m$ repetitions the computational rounds of $\mathbf{V}_n$ are used for trap computations, randomly sampled from the trappified scheme that comes with $\mathbf{V}_n$, leaving all the $m\cdot n$ sub-graphs to implement trap computations. 
        \item Instead of counting accepts vs. rejects it counts the total number of activated traps
        \item Defines small shift $\beta_t$ such that $0< \beta_t \ll \lvert \frac{\# activated}{m\cdot n} - c_t\rvert$ where $c_t$ is the threshold trap activation rate from the robust VBQC protocol
        \item Defines $z_t  = \frac{\# activated}{m\cdot n} + \beta_t$ which serves as upper bound on trap activation rate
        \item If $z_t < c_t$ the benchmarking algorithm returns $z = 1-\exp(-2n^3(c_t-z_t)^2)$ as the lower bound on the success probability
        \item If $z_t \geq c_t$ it returns $z = 0$

\end{itemize}

We call the updated benchmarking algorithm $\hat{\mathbf{B}}_{\mathbf{V}_n}$. The optimized benchmark differs merely in the classical interpretation of the data that comes back from the quantum device during the benchmark. Introducing trap computations in every computational round is very reasonable because trap computations are just one special type of possible computations and in the generic form \ref{subsec:cicc_benchmark_from_verification} these rounds are discarded. We acknowledge that the trap activation rate of a device is the quantity which governs the success probability of running the robust VBQC protocol on it. Therefore, we use every $m \cdot n$ sub-graphs to gauge this quantity.

We now prove that the optimized benchmark $(\hat{\mathbf{B}}_{\mathbf{V}_n}, \mathbf{V}_n,\mathbf{P})$ fulfills all the properties from section \ref{subsec:properties_of_a_cicc_benchmark}.

\begin{proof}[Proof of Correctness]
This proof is analogous to the generic case in section \ref{subsec:cicc_benchmark_from_verification}. When interacting with the ideal device, not a single trap will ever activate such that the returned value from the benchmark will always be $z = 1-\exp(-2n^3(c_t-\beta_t)^2)$. To see why this concludes the proof refer to the proof of correctness in \ref{subsec:cicc_benchmark_from_verification}.
\end{proof}

\begin{proof}[Proof of Correctly-Flagged Extractability]
    The proof uses similar techniques as the proof of correctly-flagged extractability in section \ref{subsec:cicc_benchmark_from_verification}. It is left out here because it does not give any further insights. Instead it can be found in the appendix \ref{sec:proof_correctly_flagged_extractability_optimized_cicc_benchmark}. The result is $\gamma$ correctly-flagged extractable with $\gamma = \exp(-2mn\beta_t^2)$
\end{proof}

\begin{proof}[Proof of Verifiable Extractability]
Verifiable extractability only considers the extracting algorithm alone which is the same as in the generic case of section \ref{subsec:cicc_benchmark_from_verification}. 
\end{proof}

\noindent \textbf{Remark.} One might wonder why we define the benchmarking algorithm to return $z=0$ if $z_t > c_t$. The reason is that if the trap activation probability is above the threshold for the scheme, then increasing the number of rounds $n$ which is needed to make the verification scheme work in the first place, diminishes the success proability as is to expected. The inherent border of computable vs. not computable for the robust VBQC protocol lies in the question whether the trap activation lies above or below the threshold of the scheme. It is the properterty that ultimately defines if $z\rightarrow 1$ or $z\rightarrow 0$ exponentially fast.

\section{A Concrete Efficient CICC Benchmarking Protocol}

In Section~\ref{sec:formal_treatment_of_benchmarking}, we have established an abstract link between quantum verification and quantum benchmarking.
In particular, we have shown that it is possible to compile a benchmarking protocol from any verification protocol with sufficiently strong security, by reinterpreting its results.
This will allow any further improvements on the side of quantum verification to carry over directly to the realm of benchmarking, through our generic compiler.

While Section~\ref{sec:formal_treatment_of_benchmarking} treated benchmarking in a very abstract way that shows conceptual advantages of reinterpreting verification as benchmarking, in this section we aim to give an explicit and self-contained description of a concrete benchmarking protocol without the abstract language of the previous section.
Security and correctness of the presented protocol directly follow from the proofs given in Section~\ref{sec:formal_treatment_of_benchmarking}.
The description of the concrete CICC benchmarking protocol based on robust VBQC protocol is given as Protocol~\ref{prot:cicc_benchmark_from_robust_vbqc}.

\begin{protocol}
    \caption{Efficient CICC Benchmark from Robust VBQC Protocol} 
    \label{prot:cicc_benchmark_from_robust_vbqc}
    \textbf{Parameters:}
    \begin{itemize}
        \item A graph $G$.
        \item A distribution of trap computations $\mathcal{D}$, and a threshold $\omega \in [0,1]$.
        \item A number of repetitions $n$.
    \end{itemize}
    \textbf{The protocol:}
    \begin{enumerate}
\item The client repeats $n$ times:
            \begin{itemize}
                \item Randomly sample a trap computation $T \sim \mathcal{D}$.
                \item Delegate the trap computation $T$ to the quantum device, using the UBQC Protocol~\ref{prot:universal_blind_quantum_computing}.
                \item Check the obtained measurement outcomes.
            \end{itemize}
            \item The client computes $z  = \frac{\#\text{failed traps}}{n}$.
            \item Iff $z < \omega$, the client returns $\textsc{Accept}$, otherwise $\textsc{Reject}$.
        \end{enumerate}        
\end{protocol}

The choice of the trap computations and the threshold will dictate the confidence in the result of the protocol and the guarantees that the benchmark can provide.
As a simple but efficient example, we refer to the generalized stabilizer traps from~\cite{kapourniotis2022unifying}. To sample from this distribution of traps, uniformly at random sample a stabilizer of the graph state $\ket{G}$ and delegate the corresponding stabilizer measurement to the quantum device, after randomization by virtue of the UBQC protocol. On a perfect and noiseless device, these measurements yield deterministic outcomes and can therefore be checked for errors. If the outcome of the stabilizer measurement is not as expected, the trap is considered to have failed.
As generalized stabilizer traps have an error detection rate of $1/2$, any threshold $\omega < 1/4$ yields a secure protocol, with the gap between $\omega$ and $1/4$ and the number of repetitions $n$ dictating the security error. For a full analysis of the security error, we refer to~\cite{kapourniotis2022unifying}.

In this sense, one can also understand this benchmarking protocol as a kind of resource state certification process with an added layer of randomization.
The resource states that we certify are graph states, and are checked for their usefulness to implement MBQC computations correctly.

\subsection{Graph State based Quantum Device Characterization}

The specific proposal for a CICC benchmark, as introduced in section \ref{CICCinstance}, was derived from formal verification protocols. It certifies computational classes $\mathcal{C}$ based on the graphs that result from translating our initial computation or unitary operation into the MBQC model. Furthermore, the deployment of our benchmark can be viewed as a process designed to characterize quantum devices in terms of the computational classes they can successfully implement. This process also involves listing these classes to provide valuable information for any future users of the devices.

\begin{figure}[ht!]
    \centering
    \begin{subfigure}[t]{0.43\textwidth} \centering
    \includegraphics[scale=0.33]{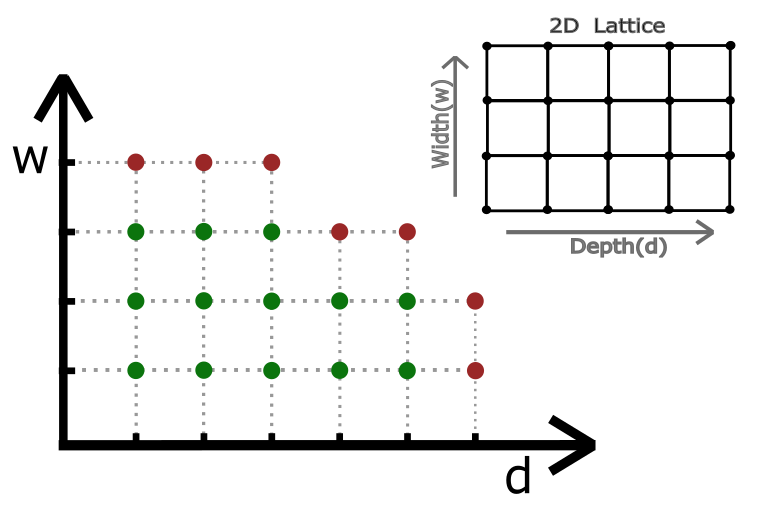} \caption{Depiction of a device characterization using tests based on graph states on 2D lattices.}
        \label{fig:sub2}
    \end{subfigure}
    \hspace{0.8cm}
    \begin{subfigure}[t]{0.50\textwidth} \centering
        \includegraphics[scale=0.40]{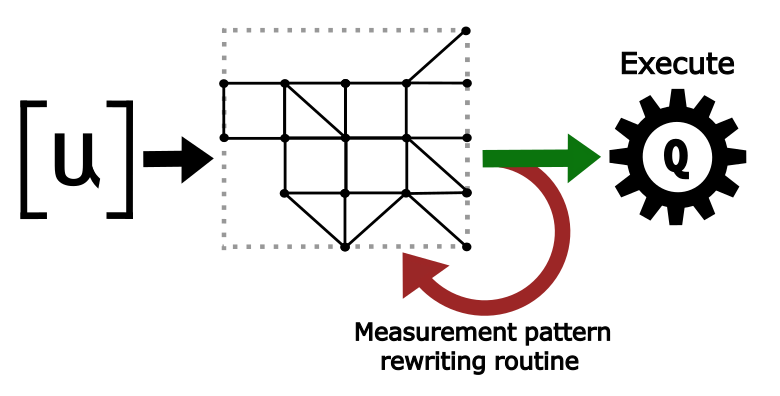} \caption{Illustration of a methodology for compiling any unitary operation onto a graph state within an MBQC pattern, accompanied by a thorough adequacy assessment to ensure accuracy.}
        \label{fig:sub1}
    \end{subfigure}
    \hfill \caption{After a quantum device is certified for different graph states, one can ask which computations, i.e., unitaries, can be certifiably run on the device. Unitaries can be compiled onto a graph state for the equivalent MBQC pattern. However, one unitary can be mapped onto different graph states. Some of these might be certified on the device, others might not. Therefore, one can attempt to rewrite the computation to find a graph state that is certified on the device so it can implement the computation.}
    \label{fig:test}
\end{figure}

In particular, as the classes under our proposal take the form of graphs, the test could be derived by benchmarking graphs of sequentially larger sizes. However, the search space does grow rapidly to an exponential size. Nevertheless, we propose, as a very informative and general characterization, the use of 2D cluster states, parameterized by two values: its width and depth. This choice results from the fact that the previous resource state is a universal resource state for the MBQC model, and the structure of the graphs resembles the characterization provided by previous benchmarks such as quantum volume, while this time providing some formal guarantees. This benchmark for quantum devices would provide a figure of merit in the form represented in Figure~\ref{fig:sub2}.

Furthermore, we would like to highlight that with such a characterization, a device user would be able to tell if its computation will be executed correctly on the device. To determine this, it is only necessary to translate the unitary into a planar MBQC measurement pattern and check if the resulting graph is equal to or a subgraph of a certified graph. Additionally, if the initial graph derived does not match any certified graph, the unitary could be rewritten according to specific rules as a deeper or more parallel computation, which might then correspond to one of the verified graphs (Figure~\ref{fig:sub1}). This further integrates the benchmark into a practical pipeline of processes that allow for optimal resource usage and computational applications of quantum devices.

\noindent \textbf{Remark.} It's important to note that the chosen graph structure is not the only possibility; the benchmark's versatility accommodates various exhaustive characterizations. Device manufacturers might tailor their device benchmarks to reflect the qubit connectivity, offering a precise characterization of quantum devices. This specificity fosters the development of circuit synthesis routines optimized for individual quantum devices, enhancing their computational performance.

\section{Discussion}

In previous verification protocols, the focus has been on establishing the correctness of specific computational instances. Our work extends this concept by eliminating the need for executing computational stages. Instead, we utilize uniquely trap computations as a generalized mechanism for characterizing quantum devices. This approach culminates in a concrete benchmarking protocol possessing all the attributes outlined in Section \ref{new_para}, filling some gaps in the existing benchmarking methods while removing the dependency on heuristics.

Practically speaking, our protocol is enunciated within the Measurement-Based Quantum Computing (MBQC) model, although translations to more commonly used circuit models can be achieved with minimal overhead \cite{simmons2021relating,Fitz18,da2013global}. This translation implies that our benchmark is hardware agnostic. Furthermore, the univerality of MBQC on 2D cluster states allows us to incorporate and enhance current benchmarking techniques by benchmarking exactly those graph states capable of implementing the computations which constitute the old benchmark.

Importantly, our benchmark focuses solely on certifying the dimensions of the employed graph states.  This allows for an exhaustive benchmark of graph state dimensions, effectively scanning the entire quantum device.  Armed with such comprehensive information, one can then synthesize circuits with the highest likelihood of successful execution on a given quantum device and obtain well-informed guidance on how the quantum solution at hand executes on a device in consideration \cite{Mio15}.

Regarding device requirements, our protocol mandates the minimal condition of supporting the interactive processes described herein. Specifically, it can be implemented in the circuit model but necessitates mid-circuit measurements. Currently, these are not universally available from all current hardware providers. However, numerous organizations are making concerted efforts to integrate this feature, attracted by its significant benefits in improving fault tolerance and aiding in circuit synthesis \cite{google2023measurement,Deist22,koh2022experimental,decross2022qubit}. Apart from this, our proposal does not necessitate a precise match between the quantum device's structure and the benchmark's specifications. Operating on a larger quantum device while consistently implementing the same processes suffices. The critical factor is ensuring appropriate communication bandwidth. We regard the remainder of the device as a "black box," thereby alleviating the need for a predefined structural framework on the quantum device, even though a specific implementation might suggest otherwise.

One interesting future question is how to reduce the vulnerability of the proposed protocol to secret-dependent noise in the communication with the quantum device, which compromises the protocol's integrity and effectiveness. Addressing verification in the presence of secret-dependent noise remains an open question and constitutes a future research avenue.
Another perceived drawback is that our benchmark appears to predominantly reject current noisy hardware. This criticism, however, stems from the non-stringent requirements of existing benchmarks rather than a flaw in our approach. Achieving more rigorous and formal guarantees inevitably comes at the cost of higher stringency, a trade-off we consider justified. Furthermore, it is worth noting that the current state of hardware has advanced to the point where the first instances of verification protocols are now realizable \cite{drmota2023verifiable,barz2013experimental}.

\paragraph*{Acknowledgements.}
This work was supported by the Quantum Advantage Pathfinder (QAP) research programme within the UK's National Quantum Computing Centre (NQCC), and by the Quantum Computing and Simulation (QCS) Hub.
JF acknowledges funding from the European Union's Horizon 2020 Research and Innovation Programme under Grant Agreement no. 731473 (QuantERA) and 101017733 (QuantERA II).
This work was supported by the European Union’s Horizon 2020 research and innovation program through the FET project PHOQUSING (“PHOtonic Quantum SamplING machine” – Grant Agreement No. 899544).
EK and DL acknowledge funding from the ANR research grant ANR-21-CE47-0014 (SecNISQ).
This work is financed by National Funds through the FCT - Fundação para a Ciência e a Tecnologia, I.P. (Portuguese Foundation for Science and Technology) within the project IBEX, with reference PTDC/CCI-COM/4280/2021, and via CEECINST/00062/2018 (EFG).

\bibliographystyle{splncs04}
\bibliography{biblio.bib}

\clearpage
\appendix

\section{Hoeffding's Inequality}
\label{sec:hoeffding}
We use the Hoeffding's inequality for $X_1, \dots , X_m$ independent random variables with $a_i \leq X_i \leq b_i$. Consider $S_m = X_1 + \dots + X_m$. Then it holds:
\begin{equation}
    \forall t>0: \Pr[S_m - \mathbb{E}[S_m]\geq t] \leq \exp(-\frac{2t^2}{\sum_{i=1}^{m}(b_i-a_i)^2})
\end{equation}
If we substitute $t = m\Tilde{t}$ we obtain
\begin{equation}
    \Pr[\frac{S_m}{m} - \frac{\mathbb{E}[S_m]}{m} \geq \Tilde{t}] \leq \exp(-\frac{2m^2\Tilde{t}^2}{\sum_{i=1}^{m}(b_i-a_i)^2}),
\end{equation}
because $b_i = 1$ and $a_i = 0$. Now we utilize the fact that $X_i$ are i.i.d. binary random variables. Therefore 
\begin{equation}
    \mathbb{E}[S_m] = m\mathbb{E}[X] = mp
\end{equation}
where $p$ is the probability of getting 1 as result. Also: $(b_i - a_i)^2 = 1$. At the same time we rename $\frac{S_m}{m} = z$ to obtain 
\begin{equation}
    \Pr[z-p\geq t] \leq \exp(-2mt^2).
\end{equation}
Finally, rearranging gives us
\begin{equation}
    \Pr[p\leq z-t]\leq \exp(-2mt^2),
\end{equation}
which is the Hoeffding's inequality for $m$ repeated biased coin flips.

\section{Proof of Correctly-Flagged Extractability of the Optimized CICC 
Benchmark using Robust VBQC}

\label{sec:proof_correctly_flagged_extractability_optimized_cicc_benchmark}

\begin{proof}[Proof of Correctly-Flagged Extractability]

We first note that the the probability of a trap activating is the same on all of the $m\cdot n$ sub-graph trap computations. Why? By definition, the behavior of the device is unchanged for all $m$ repetitions of the all-trappified robust VBQC protocol. In section \ref{subsec:recap_robust_vbqc_protocol} we also explained that for a robust VBQC protocol $\mathbf{V}_n$ the global error model, that defines probability of errors on $G^n$, can be reduced to a local error model on the sub-graph $G$ that is sampled $n$ times. Effectively, we therefore have $m\cdot n$ independent trials of randomly sampling both trap and error for each sub-graph state $G$. Every of these $m\cdot n$ rounds will have the same probability of activating a trap. This probability $p_t$ is unknown upfront. However, we are again in the situation of $m \cdot n$ Bernoulli trials.

We define $z_t  = \frac{\# activated}{m\cdot n} + \beta_t$ for some $0<\beta_t \ll \lvert \frac{\# activated}{m\cdot n} - c_t\rvert$. Hoeffding's inequality yields for trap activation probability $p_t$ in each round: 

\begin{equation}
\label{eq:hoeffdings_inequality_trap_activation_probability}
    \Pr[p_t < z_t] > 1- \exp(-2mn\beta_t^2)
\end{equation}
Now we want to use \ref{eq:hoeffdings_inequality_trap_activation_probability} to find out the acceptance probability of $\mathbf{V}_n$. Whether the robust VBQC accepts or not is determined by the number of activated traps. We recognize the execution of $\langle \mathbf{V}_n,\mathbf{Q}\rangle$ again as an $n$ round Bernoulli trial with random variables $X_i$:
\begin{equation}
    \forall i \in [n]: X_i \sim \begin{cases}
                                    1 & \text{with probability } p_t \\
                                    0 & \text{with probability } 1 - p_t
                                \end{cases}
\end{equation}
where we identify outcome 1 with and activated trap and 0 for not activating a trap for each round. We denote the precise condition under which the verification scheme accepts:
\begin{equation}
    S_n = \sum_{i=1}^n X_i < c_t \cdot n
\end{equation}
Where the exact value of $c_t$ is defined in the specific choice of the scheme. Refer to \ref{subsec:recap_robust_vbqc_protocol}.

Just like in the definition of $\hat{\mathbf{B}}_{\mathbf{V}_n}$ we now make a case distinction: \newline

\noindent \textbf{Case 1: $z_t < c_t$}
From \ref{eq:hoeffdings_inequality_trap_activation_probability} we know that with probability $p_t <z_t< c_t$ which is what assume for the following argument. We use Hoeffdings inequality: 
\begin{equation}
   \Pr[S_n \geq \mathbb{E}[S_n] + t] \leq \exp(-2nt^2),  t > 0
\end{equation}
We note that $\mathbb{E}[S_n] = np_t$ and set $t = n(c_t - p_t) > 0$ to obtain:
\begin{equation}
    \Pr[S_n \geq nc_t] \leq \exp(-2n^3(c_t - p_t)^2)
\end{equation}
However, $S_n \geq nc_t$ is exactly the condition under which $\mathbf{V}_n$ \textbf{rejects} the result. We conclude: 
\begin{equation}
    \Pr[\mathbf{V}_n \; \mathrm{accepts}] > 1- \exp(-2n^3(c_t - p_t)^2) \overset{z_t > p_t}{>} 1- \exp(-2n^3(c_t - z_t)^2) 
\end{equation}

Now we remember that the assumption $p_t < z_t $ was actually only justified with high probability (\ref{eq:hoeffdings_inequality_trap_activation_probability}). We get:
\begin{equation}
    \Pr[\Pr[\mathbf{V}_n  \; \mathrm{accepts}] > 1-  \exp(-2n^3(c_t - z_t)^2)]  > 1 - \exp(-2mn\beta_t^2)
\end{equation}
which is equivalent to : 

\begin{equation}
    \Pr[\Pr[\mathbf{V}_n \; \mathrm{accepts}] \leq  1 - \exp(-2n^3(c_t - z_t)^2)] \leq \exp(-2mn\beta_t^2)
\end{equation}
We remind ourselves of the output of $\hat{\mathbf{B}}_{\mathbf{V}_n}$: $z =  1 - \exp(-2n^3(c_t - z_t)^2)$. We can directly identify correctly-flagged extractability from this expression in the case of $z_t < c_t$.\newline

\noindent \textbf{Case 2: $z_t \geq c_t$}
The benchmark simply returns $z=0$. It trivially holds: 

\begin{equation}
    \Pr[\Pr[\mathbf{V}_n \; \mathrm{accepts}] \leq 0] = 0 < \exp(-2mn\beta_t^2)
\end{equation}

We combine \textbf{Case 1} and \textbf{Case 2} to conclude that for any value $z_t$, and therefore in general, the CICC benchmark defined by $(\hat{\mathbf{B}}_{\mathbf{V}_n}, \mathbf{V}_n, \mathbf{P})$ is $\gamma$ correctly-flagged extractable with $\gamma = \exp(-2mn\beta_t^2)$

\end{proof}

\section{Measurement Based Quantum Computing and Universal Blind Quantum Computation}
\label{subsec:mbqc_ubqc_robust_vbqc}

This introduction mainly follows \cite{kapourniotis2022unifying}. Measurement based quantum computing (MBQC)
utilizes gate teleportation to implement universal quantum computations based on graph-states as entanglement resource and single qubit measurements \cite{PhysRevLett.86.5188}. Graph states from graph $G = (V,E)$ where $V$ is the set of vertices or nodes and $E$ is the set of edges between them are defined as follows:

\begin{equation}
\label{eq:graph_state_definition}
\left|G\right\rangle = \prod_{(a,b)\in E} CZ(a,b) \left|+\right\rangle^{\otimes V}
\end{equation}

While it is perfectly possible to use quantum input and output computations with this model, in this work we restrict to the classical input/output case where we can carry out MBQC in a delegation scheme where the client only communicates the measurement angles to a large quantum server that can operate a graph state with large scale entanglement.

Computations are done in the MBQC picture as so-called measurement patterns. That means we choose a graph $G= (V,E)$ and a set of measurement angles $\{ \phi(i)\}_{i \in V}$ for all the qubits in the graph state $\left|G\right\rangle$ and a flow function $f$ which defines a partial order on the qubits of $\left|G\right\rangle$. Carrying out a computation then means measuring according to the flow $f$ each qubit $i$ of $\left|G\right\rangle$ in the X-Y plane according to the (corrected) measurement angles $\phi(i)$. Depending on the measurement outcome of each qubit, the computation will be carried out with a different set of measurement angles $\phi'(i)$ that are adaptively computed using the classical information of previous measurement outcomes we keep track of. That means that the initial set $\{ \phi(i)\}_{i \in V}$ is only used if all the intermediate measurement results are zero.

MBQC can be realized in a delegation setting where we have a client with small quantum capacities i.e. the client is able to prepare and send single qubits states from a finite set. Then the server has universal quantum computational capacities and carries out the entangling operations needed for producing the graph states. Then the client communicates layer by layer (graph states can often be measured out layer-wise i.e. multiple qubits are measured simultaneously because their outcomes do not influence the measurement angles of each other) the measurement angles. Then the server carries out the measurement in the correct basis and communicates the outcome. Then the client updates the angles for the next layer until the entire graph state is measured out. The last layer will typically be the outcome of the computation.

Unlike one might assume from the graph state construction MBQC can be done with less operational qubits than constitute the graph state. Early qubits and later qubits do not necessarily be stored in quantum memory at the same time, but can be provided "on the fly" \cite{Brien07}.

We provide the protocol of delegated measurement based quantum computing as the benchmarking protocol that we propose will be based on a slightly adapted version. We use the notation of \cite{kapourniotis2022unifying} and particularly, 
\begin{equation}
\left|\pm_{\theta}\right\rangle = \frac{1}{\sqrt{2}}(\left|0\right\rangle \pm e^{i\theta} \left|1\right\rangle).
\end{equation}.

\begin{protocol}
    \caption{Measurement Based Quantum Computing} 
    \label{prot:measurement_based_quantum_computing}
        \begin{enumerate}
            \item Client sends Graph $G = (V,E)$, flow $f$ and $S_X(j)$ and $S_Z(j)$, the subsets of qubits that influence the angle of qubit $j$, derived from $G$ and $f$ to the server
            \item Server prepares $\left|G\right\rangle$
            \item Client sends measurement angles $\{ \phi(i)\}_{i \in V}$ to server which encode the target computation
            \item Server measures qubits in the order induced by $f$ in the basis $\left|\pm_{\phi'(i)}\right\rangle$. $\phi'(i)$ is computed by the server as 
            \begin{equation}
                \phi'(i) =  (-1)^{s_X(i)}\phi(i) + s_Z(i)\pi 
            \end{equation}
            \begin{equation}
                s_{X}(i) = \mathop{\bigoplus}\limits_{j \in S_{X}(i)} b(j), \quad s_{Z}(i) = \mathop{\bigoplus}\limits_{j \in S_{Z}(i)} b(j)
            \end{equation}
            where $b(j)$ is the measurement outcome of qubit $j$. 
            \item Server communicates last measurement layer to the client as result of the target computation
        \end{enumerate}
\end{protocol}

The delegated version of MBQC that we discussed here allows for a client with limited quantum resources i.e. preparing and sending single qubit states to delegate classical input and output computations to a computationally more capable server.

However, in this way the server gets full access to the classical description of the data and algorithm involved in the computation. Especially in a cloud setting this is highly undesirable.
Instead, what a client can do is to hide all of the information that is used in protocol \ref{prot:measurement_based_quantum_computing} except for the graph $G$ and flow $f$, which merely disclose the upper limit of resources employed by the server, by completely randomizing the communicated angles. We provide here the universal blind computation protocol from \cite{broadbent2009universal} but do not prove its desirable properties of composable security for delegating quantum computations in a blind way. We describe the protocol rather than explain it here. For a more rigorous understanding of the protocol the reader is referred to \cite{broadbent2009universal}. Again we use the notation of \cite{kapourniotis2022unifying}.

\begin{protocol}
    \caption{Universal Blind Quantum Computing} 
    \label{prot:universal_blind_quantum_computing}
        \begin{enumerate}
            \item Client sends Graph $G = (V,E)$ and flow $f$ to the server
            \item Client prepares and sends all the qubits $i$ in $V$ as either 
            \begin{itemize}
                \item $\left|+_{\theta(i)}\right\rangle$ where $\theta(i)$ is randomly sampled for every qubit $i\in V$ from $\theta = \{0, \frac{\pi}{4}, \dots , \frac{7\pi}{4}\}$
                \item or $X^{r(i)}\ket{0}$ where $r(i)$ is randomly sampled from $\{0,1\}$. For X-Y basis qubits $r(i) = 0$ per default.
            \end{itemize}
            
            \item Server entangles all the qubits just as in protocol \ref{prot:measurement_based_quantum_computing} $\left|G\right\rangle$
            \item Client sends measurement angles $\{ \delta(i)\}_{i \in V}$ to server for every measurement, right before the measurement. $\delta(i)$ is computed by the client as:
            \begin{equation}
                \delta(i) =  (-1)^{s_X(i)}\phi(i) + s_Z(i)\pi+ \theta(i) +r(i)\pi
            \end{equation}
            \begin{equation}
                s_{X}(i) = \mathop{\bigoplus}\limits_{j \in S_{X}(i)} b(j), \quad s_{Z}(i) = \mathop{\bigoplus}\limits_{j \in S_{Z}(i)} b(j)
            \end{equation}
            
            where $b(j)$ is the measurement outcome of qubit $j$. 
            \item Server measures qubits in the basis $\left|\pm_{\delta(i)}\right\rangle$ and communicates the result directly to the client who uses the information for the next measurement layer.            
            \item Server communicates last measurement layer to the client as result of the target computation.
        \end{enumerate}
\end{protocol} 

\end{document}